# Scattering phase shift for relativistic separable potential with Laguerre-type form factors

**A. D. Alhaidari**
*Physics Department, King Fahd University of Petroleum & Minerals, Box 5047, Dhahran 31261, Saudi Arabia*
E-mail: **haidari@mailaps.org**

As an extension of earlier work [J. Phys. A: *Math. Gen.* **34** (2001) 11273] we obtain analytic expressions for the scattering phase shift of *M*-term relativistic separable potential with Laguerre-type form factors and for $M$ = 1, 2, and 3. We take the Dirac Hamiltonian as the reference Hamiltonian. Just like in the cited article, the tools of the relativistic J-matrix method of scattering will be used. However, the results obtained here are for a general angular momentum, which is in contrast to the previous work where only S-wave scattering could be calculated. An exact numerical evaluation for higher order potentials ($M \geq 4$) can be obtained in a simple and straightforward way.

In a previous article [1], we considered relativistic scattering for separable potentials with exponential–type form factors and took the Dirac Hamiltonian as reference Hamiltonian. We investigated two classes of these potentials whose radial component is either of the general form $r^{\nu-1}e^{-\lambda r/2}$ or $r^{2\nu}e^{-\lambda^2 r^2/2}$, where $\lambda$ is a range parameter and $\nu$ = 0, 1, 2, and 3. It turned out that these potentials are ($\nu$ + 1)–term separable, that is their matrix representations in a suitable $L^2$ basis are of dimension $\nu$ + 1. The tools of the relativistic J–matrix method of scattering [2–4] were used to obtain analytic expressions for the phase shift for $\nu$ = 0, 1, and 2. For higher order potentials ($\nu \geq 3$), we found that it was sufficient and more practical to calculate the exact phase shift numerically using the relativistic J–matrix method with a function space dimension $N \geq \nu + 1$. The resulting structure of the phase shift as a function of energy turned out to be very rich and highly interesting as demonstrated graphically in [1]. In this short article we extend that work significantly and in two directions. On one hand, the form factors in the relativistic separable potential will be Laguerre–type of the form $(\lambda r)^{l+1}e^{-x/2}L_n^{\mu}(x)$ where $L_n^{\mu}(x)$ is the generalized Laguerre polynomial [5] and either $x = \lambda r$ or $x = (\lambda r)^2$. On the other hand, the results obtained here are for any value of the angular momentum which is in contrast to the previous work where we could only obtain the phase shift for S–wave scattering. We start by setting up the problem but the method of solution used, which can be found in [1], will not be repeated here.

In atomic units ($m = e = \hbar = 1$) and taking the speed of light $c = \alpha^{-1}$, the free Dirac equation reads

$$\begin{pmatrix} 1-\varepsilon & \alpha\left(\frac{\kappa}{r}-\frac{d}{dr}\right) \\ \alpha\left(\frac{\kappa}{r}+\frac{d}{dr}\right) & -1-\varepsilon \end{pmatrix} \begin{pmatrix} \sum_n h_n(\varepsilon)\phi_n(r) \\ \sum_n h_n(\varepsilon)\theta_n(r) \end{pmatrix} = 0 \qquad (1)$$

where $\alpha$ is the fine structure constant, $\kappa$ is the spin–orbit coupling parameter defined by $\kappa = \pm\,(j + ½)$ for $l = j \pm ½$, $\varepsilon$ is the relativistic energy, and $\{h_n(\varepsilon)\}_{n=0}^{\infty}$ is the set of expansion coefficients of the spinor wave function. We let $\psi_n(r)$ stand for the two–component radial spinor basis whose upper component is $\phi_n(r)$ and lower component $\theta_n(r)$ and consider the following $M$–term separable potential

$$\tilde{V} = \sum_{n,m=0}^{M-1} \left|\overline{\psi}_n\right\rangle V_{nm} \left\langle\overline{\psi}_m\right|$$

where $\overline{\psi}_n$ is an element of the conjugate $L^2$ space [2] (i.e., $\langle\overline{\psi}_n|\psi_m\rangle = \langle\psi_n|\overline{\psi}_m\rangle = \delta_{nm}$) and $V$ is an $M\times M$ real symmetric matrix. In configuration space this separable potential can be written as

$$\tilde{V}(r,r') \equiv \langle r|\tilde{V}|r'\rangle = \begin{pmatrix} V^+(r,r') & V^0(r,r') \\ V^0(r',r) & V^-(r,r') \end{pmatrix}$$

where

$$V^+(r,r') = \sum_{n,m=0}^{M-1} V_{nm}\overline{\phi}_n(r)\overline{\phi}_m(r')$$

$$V^-(r,r') = \sum_{n,m=0}^{M-1} V_{nm}\overline{\theta}_n(r)\overline{\theta}_m(r')$$

$$V^0(r,r') = \sum_{n,m=0}^{M-1} V_{nm}\overline{\phi}_n(r)\overline{\theta}_m(r')$$

Including the separable potential in the wave equation (1) gives

$$\sum_{m=0}^{\infty} h_m\left(H_0 - \varepsilon\right)|\psi_m\rangle + \alpha^2 \sum_{m,k=0}^{M-1} V_{km} h_m |\overline{\psi}_k\rangle = 0 \qquad (2)$$

where $H_0$ is the reference Hamiltonian, that is the 2×2 matrix operator in (1) with $\varepsilon = 0$. Projecting equation (2) on $\langle\psi_n|$, we obtain the following equivalent matrix wave equation:

$$\sum_{m=0}^{\infty} \Im_{nm}(\varepsilon) h_m(\varepsilon) + \alpha^2 \sum_{m=0}^{M-1} V_{nm} h_m(\varepsilon) = 0 \qquad ; n = 0,1,2,\ldots \qquad (3)$$

where $\Im_{nm}(\varepsilon) \equiv \left(H_0\right)_{nm} - \varepsilon\, \Omega_{nm}$, and $\Omega$ is the matrix representation of the identity, i.e. the basis overlap matrix whose elements are $\Omega_{nm} = \langle\psi_n|\psi_m\rangle$. In the J–matrix method, the bases $\{\psi_n\}_{n=0}^{\infty}$ are chosen such that the matrix representations of $H_0$ and $\Omega$ are tridiagonal, thus so is $\Im$. Therefore, equation (3) could be written explicitly in the following matrix form

$$\left[\begin{pmatrix} \Im_{00} & \Im_{01} & & & & \\ \Im_{10} & \Im_{11} & \Im_{12} & & 0 & \\ & \Im_{21} & \Im_{22} & \Im_{23} & & \\ & & .. & .. & .. & \\ & 0 & & .. & .. & .. \\ & & & & .. & .. \end{pmatrix} + \alpha^2 \begin{pmatrix} V_{00} & V_{01} & .. & V_{0,M-1} & 0 & .. \\ V_{10} & V_{11} & .. & V_{1,M-1} & 0 & .. \\ \vdots & \vdots & .. & \vdots & \vdots & .. \\ V_{M-1,0} & V_{M-1,1} & .. & V_{M-1,M-1} & 0 & .. \\ 0 & 0 & .. & 0 & 0 & .. \\ \vdots & \vdots & .. & \vdots & \vdots & .. \end{pmatrix}\right] \begin{pmatrix} h_0 \\ h_1 \\ \vdots \\ h_{M-1} \\ h_M \\ \vdots \end{pmatrix} = 0 \qquad (4)$$

This relativistic J–matrix scattering problem has an *exact* numerical solution which is obtained by taking the $M\times M$ matrix representation of the separable potential as the exact short–range perturbing potential in the standard J–matrix scheme. To obtain this solution, we proceed as follows:

For an integer $N \geq M$ the $M\times M$ potential matrix, $\left\{\alpha^2 V_{nm}\right\}_{n,m=0}^{M-1}$, will be added to the $N\times N$ tridiagonal matrix representation of the reference Hamiltonian (listed in the Table for the two types of separable potentials in the two corresponding bases) giving the $N\times N$ total Hamiltonian. This is then diagonalized and used in the calculation of the finite Green's function leading to the scattering phase shift in the usual way [2,4,6]. However, it turns out that without too much difficulty and with a reasonable effort we can obtain an *analytic* solution for the relativistic phase shift for $M$ = 1, 2, and 3. This is accomplished by considering a *new* relativistic J–matrix problem in which the reference Hamiltonian, $\hat{H}_0$, is the sum of the tridiagonal kinetic energy term, $H_0$, and this separable potential. It is evident from the matrix wave equation (4) that the resulting symmetric recursion relation for the expansion coefficients of the wavefunction is asymptotically ($n \geq M$) three–term. While, the first set of $M$ equations for $\left\{h_n(\varepsilon)\right\}_{n=0}^{M-1}$ can be thought of as the initial relations for this three–term recursion. That is, the separable potential acts like a source term for the recursion relation. The analytic solution of the recursion relation, with these newly emerging initial conditions due to the separable potential, gives the new sine–like, $h_n = \hat{s}_n$, and cosine–like, $h_n = \hat{c}_n$, expansion coefficients of the wave function. The asymptotic behavior of these expansion coefficients gives the sought after phase shift as a rotation angle of the original coefficients $\left\{s_n, c_n\right\}_{n\geq M-1}$. To obtain a more compact and transparent solution we write the problem in terms of the complex coefficients defined by

$$g_n^{\pm}(\varepsilon) = c_n(\varepsilon) \pm i s_n(\varepsilon)$$

In this notation, the asymptotic form of equation (4) gives the following symmetric three–term recursion relation

$$\Im_{n,n-1}\hat{g}_{n-1}^{+} + \Im_{n,n}\hat{g}_n^{+} + \Im_{n,n+1}\hat{g}_{n+1}^{+} = 0 \quad ; \quad n \geq M$$

which is the same as that of the original J–matrix problem for $n \geq 1$ [1,2,6]. The initial relations, on the other hand, are as follows:

$$\begin{pmatrix} \Im_{00} & \Im_{01} & & & & & 0 \\ \Im_{10} & \Im_{11} & \Im_{12} & & & & \\ & \Im_{21} & \Im_{22} & \Im_{23} & & & \\ & & \cdot\cdot & \cdot\cdot & \cdot\cdot & & \\ 0 & & & & \Im_{M-1,M-2} & \Im_{M-1,M-1} & \Im_{M-1,M} \end{pmatrix} \begin{pmatrix} \hat{g}_0^{+} \\ \hat{g}_1^{+} \\ \hat{g}_2^{+} \\ \vdots \\ \hat{g}_{M-1}^{+} \\ \hat{g}_M^{+} \end{pmatrix} = -\alpha^2 \begin{pmatrix} V_{00} & V_{01} & \cdot\cdot & \cdot\cdot & V_{0,M-1} \\ V_{10} & V_{11} & \cdot\cdot & \cdot\cdot & V_{1,M-1} \\ \vdots & \vdots & \cdot\cdot & \cdot\cdot & \vdots \\ \vdots & \vdots & \cdot\cdot & \cdot\cdot & \vdots \\ V_{M-1,0} & V_{M-1,1} & \cdot\cdot & \cdot\cdot & V_{M-1,M-1} \end{pmatrix} \begin{pmatrix} \hat{g}_0^{+} \\ \hat{g}_1^{+} \\ \vdots \\ \vdots \\ \hat{g}_{M-1}^{+} \end{pmatrix} - \frac{i\alpha^2 w/\hat{g}_0^{+}}{1-\hat{T}_0} \begin{pmatrix} 1 \\ 0 \\ \vdots \\ \vdots \\ 0 \end{pmatrix}$$

where $w(\varepsilon)$ is the Wronskian of the regular and irregular solutions of the free Dirac problem [2–4].

To recover the original recursion, whose initial relation is

$$\Im_{00}g_0^{+} + \Im_{01}g_1^{+} = -i\alpha^2 w/g_0^{+}(1-T_0)$$

we propose the following transformation of the complex coefficients:

$$\begin{aligned} \hat{g}_m^{\pm} &= \mu_m e^{\pm i\sigma_m} g_m^{\pm} \quad ; \quad m = 0,1,..,M-2 \\ \hat{g}_n^{\pm} &= e^{\pm i\tau} g_n^{\pm} \quad ; n \geq M-1 \end{aligned} \tag{5}$$

where $\{\mu_m\}$ and $\{\sigma_m\}$ are real constant parameters, and $\mu_m > 0$. The first line in (5) is a $2(M-1)$-parameter scaling and rotation transformation. However, the last line, which shows the asymptotic behavior of the expansion coefficients, could be interpreted to give the phase shift due to the separable potential as the rotation angle, $\tau$, of the original coefficients.

Following the same computational steps as in reference [1] we arrive at the analytic expressions giving the phase shift angle $\tau$ for the first three $M$–term separable potentials. In these expressions, which are given below, the J–matrix kinematical coefficients $T_n(\varepsilon)$ and $R_n^{\pm}(\varepsilon)$ are defined as

$$T_n \equiv \frac{c_n - is_n}{c_n + is_n} = \frac{g_n^-}{g_n^+} \quad ; \quad R_{n+1}^{\pm} \equiv \frac{c_{n+1} \pm is_{n+1}}{c_n \pm is_n} = \frac{g_{n+1}^{\pm}}{g_n^{\pm}} \qquad ; n \geq 0$$

The wavefunction expansion coefficients $\{s_n(\varepsilon)\}$ and $\{c_n(\varepsilon)\}$ for the $H_0$–problem in the Laguerre and oscillator basis are given in the Table. The tridiagonal matrix elements of $H_0$ and $\Omega$ needed to calculate $\Im$ for the two types of separable potentials are also listed in the Table.

***M* = 1:**

$$e^{2i\tau} = T_0 + \left(1 - T_0\right)\left[1 + \frac{\alpha^2 V_{00}}{\Im_{00} + \Im_{01} R_1^+}\right]^{-1}$$

***M* = 2:**

$$e^{2i\tau} = T_0 e^{-2i\zeta} + \left(1 - T_0\right)\left(\Im_{01} + \alpha^2 V_{01}\right)\left(\frac{\Im_{00} + \Im_{01} R_1^+}{\Im_{01} - \alpha^2 V_{11} R_1^+}\right) \times$$
$$\left[\left(\Im_{01} - \alpha^2 V_{11} R_1^+\right)\frac{\Im_{00} + \alpha^2 V_{00}}{\Im_{01} + \alpha^2 V_{01}} + R_1^+\left(\Im_{01} + \alpha^2 V_{01}\right)\right]^{-1}$$

where $\zeta = \arg\left[\left(\Im_{01} - \alpha^2 V_{11} R_1^+\right) \big/ \left(\Im_{01} + \alpha^2 V_{01}\right)\right]$.

***M* = 3:**

$$e^{2i\tau} = T_0 e^{-2i\xi} + \frac{1 - T_0}{R_1^+ \Lambda}\left(\Im_{00} + R_1^+ \Im_{01}\right) \times$$
$$\left\{R_1^+ \Lambda\left(\Im_{00} + \alpha^2 V_{00}\right) + \alpha^2 R_1^+\left[\frac{\Im_{01} + \alpha^2 V_{01}}{\Im_{12} + \alpha^2 V_{12}}\left(\Im_{12}/\alpha^2 - R_2^+ V_{22} - V_{02}\Lambda\right) + R_2^+ V_{02}\right]\right\}^{-1}$$

where $\xi = \arg\left(R_1^+ \Lambda\right)$ and,

$$\Lambda(\varepsilon, V) = \frac{\left(\Im_{01}/R_1^+\right) + \Im_{11} - \alpha^2 R_2^+ V_{12} + \dfrac{\Im_{11} + \alpha^2 V_{11}}{\Im_{12} + \alpha^2 V_{12}}\left(-\Im_{12} + \alpha^2 R_2^+ V_{22}\right)}{\Im_{01} + \alpha^2 V_{01} - \alpha^2 V_{02}\dfrac{\Im_{11} + \alpha^2 V_{11}}{\Im_{12} + \alpha^2 V_{12}}}$$

The nonrelativistic limit of these expressions is obtained by taking $\alpha \to 0$ and $C = \alpha/2$ [1,2,4], where $C$ is the strength parameter for the small spinor component which also appears in the Table.

The analytic expressions above for the scattering matrix $e^{2i\tau}$ are almost identical to those in [1]. However, note that $\{V_{nm}\}$ are the separable potential parameters given as input to the problem while in [1] they were evaluated as matrix elements of the exponential potentials by integration in the $L^2$ basis. Moreover, when calculating $\Im_{nm}$, $T_n$ and $R_n^\pm$ in the formulas above we should use their expressions for a general angular momentum as given in the Table not those in reference [1] for $l = 0$.

## TABLE CAPTION

For the first and second type of separable potentials the following quantities are listed in the Laguerre and oscillator spinor basis, respectively: the upper component of the basis, the tridiagonal matrix elements of $H_0$ and $\Omega$, the sine–like and cosine–like coefficients $\{s_n(\varepsilon)\}$ and $\{c_n(\varepsilon)\}$ [2,6]. In the table ${}_2F_1(a,b;c;z)$ is the hypergeometric function, ${}_1F_1(a;b;z)$ is the confluent hypergeometric function, $C_n^{\nu}(x)$ is the Gegenbauer polynomial, and $\eta(\varepsilon) \equiv K(\varepsilon)/\lambda$, where

$$K(\varepsilon) = \sqrt{\frac{-1}{C^2}\frac{\varepsilon - 1}{\varepsilon - 1 + 2(1 - \alpha / C)}}$$

The angle $\omega(\varepsilon)$ is defined by:

$$\cos(\omega) = \frac{\left[K(\varepsilon)/\lambda\right]^2 - 1/4}{\left[K(\varepsilon)/\lambda\right]^2 + 1/4}$$

**TABLE**

| | Laguerre basis | Oscillator basis |
|---|---|---|
| $\phi_n(r)$ | $a_n(\lambda r)^{\kappa+1} e^{-\lambda r/2} L_n^{2\kappa+1}(\lambda r)$<br>$a_n = \sqrt{\lambda\Gamma(n+1)/\Gamma(2\kappa+n+2)}$ | $a_n(\lambda r)^{\kappa+1} e^{-\lambda^2 r^2/2} L_n^{\kappa+1/2}(\lambda^2 r^2)$<br>$a_n = \sqrt{2\lambda\Gamma(n+1)/\Gamma(n+\kappa+3/2)}$ |
| $(H_0)_{nm}$ | $(H_0)_{n,n} = 2(\kappa+n+1)\left[1-(\lambda C/2)^2(1-2\alpha/C)\right]$<br>$(H_0)_{n,n-1} = -\sqrt{n(2\kappa+n+1)}\left[1+(\lambda C/2)^2(1-2\alpha/C)\right]$<br>$(H_0)_{n,n+1} = -\sqrt{(n+1)(2\kappa+n+2)}\left[1+(\lambda C/2)^2(1-2\alpha/C)\right]$ | $(H_0)_{nn} = 1+\lambda^2 C^2(-1+2\alpha/C)(2n+\kappa+3/2)$<br>$(H_0)_{n,n-1} = \lambda^2 C^2(-1+2\alpha/C)\sqrt{n(n+\kappa+1/2)}$<br>$(H_0)_{n,n+1} = \lambda^2 C^2(-1+2\alpha/C)\sqrt{(n+1)(n+\kappa+3/2)}$ |
| $\Omega_{nm}$ | $\Omega_{n,n} = 2(\kappa+n+1)\left[1+(\lambda C/2)^2\right]$<br>$\Omega_{n,n-1} = -\sqrt{n(2\kappa+n+1)}\left[1-(\lambda C/2)^2\right]$<br>$\Omega_{n,n+1} = -\sqrt{(n+1)(2\kappa+n+2)}\left[1-(\lambda C/2)^2\right]$ | $\Omega_{nn} = 1+\lambda^2 C^2(2n+\kappa+3/2)$<br>$\Omega_{n,n-1} = \lambda^2 C^2\sqrt{n(n+\kappa+1/2)}$<br>$\Omega_{n,n+1} = \lambda^2 C^2\sqrt{(n+1)(n+\kappa+3/2)}$ |
| $c_n(\varepsilon)$ | $-\frac{2^\kappa}{\sqrt{\pi}}\frac{a_n}{\lambda}\frac{\Gamma(\kappa+1/2)}{(\sin\omega)^\kappa}\times$<br>${}_2F_1\left(-n-1-2\kappa, n+1; 1/2-\kappa; \sin^2\frac{\omega}{2}\right)$ | $(-1)^n\frac{\Gamma(\kappa+1/2)}{\sqrt{2\pi}}\frac{a_n}{\lambda}\eta^{-\kappa}e^{-\eta^2/2}\times$<br>${}_1F_1\left(-n-1/2-\kappa; 1/2-\kappa; \eta^2\right)$ |
| $s_n(\varepsilon)$ | $\frac{2^\kappa}{\lambda}a_n\Gamma(\kappa+1)(\sin\omega)^{\kappa+1}C_n^{\kappa+1}(\cos\omega)$ | $\frac{(-1)^n}{\lambda}\sqrt{\frac{\pi}{2}}a_n\eta^{\kappa+1}e^{-\eta^2/2}L_n^{\kappa+1/2}(\eta^2)$ |